\documentclass[twocolumn]{article}
\usepackage{graphicx} %
\usepackage{amsmath}
\usepackage[percent]{overpic}
\usepackage{color}
\usepackage{hyperref}

\title{Summary of the 16th Applied Antineutrino Physics Workshop
2023}
\author{Compiled by Liz Kneale and Viacheslav Li\footnote{Lawrence Livermore National Laboratory is operated by Lawrence Livermore National Security, LLC, for the U.S. Department of Energy, National Nuclear Security Administration under Contract DE-AC52-07NA27344. LLNL-PROC-871170.}}

\begin{document}

\maketitle

\section*{Day 1: Overview lectures and neutrino-application reports}

The 16$^{\mathrm{th}}$ Applied Antineutrino Physics (AAP) Workshop was held in the historic Guildhall in York, UK on September 18--21, 2023. 
This was the first workshop after the COVID-19 pandemic. 
The 3-day workshop was followed by a tour at the STFC Boulby Underground Laboratory.
A conference dinner was held %
at the 14$^{\mathrm{th}}$-century Hospitium, set in the beautiful York Museum Gardens,  overlooked by the striking ruins of St. Mary’s Abbey.

\begin{figure*}[ht]
    \centering
    \includegraphics[width=1.\textwidth]{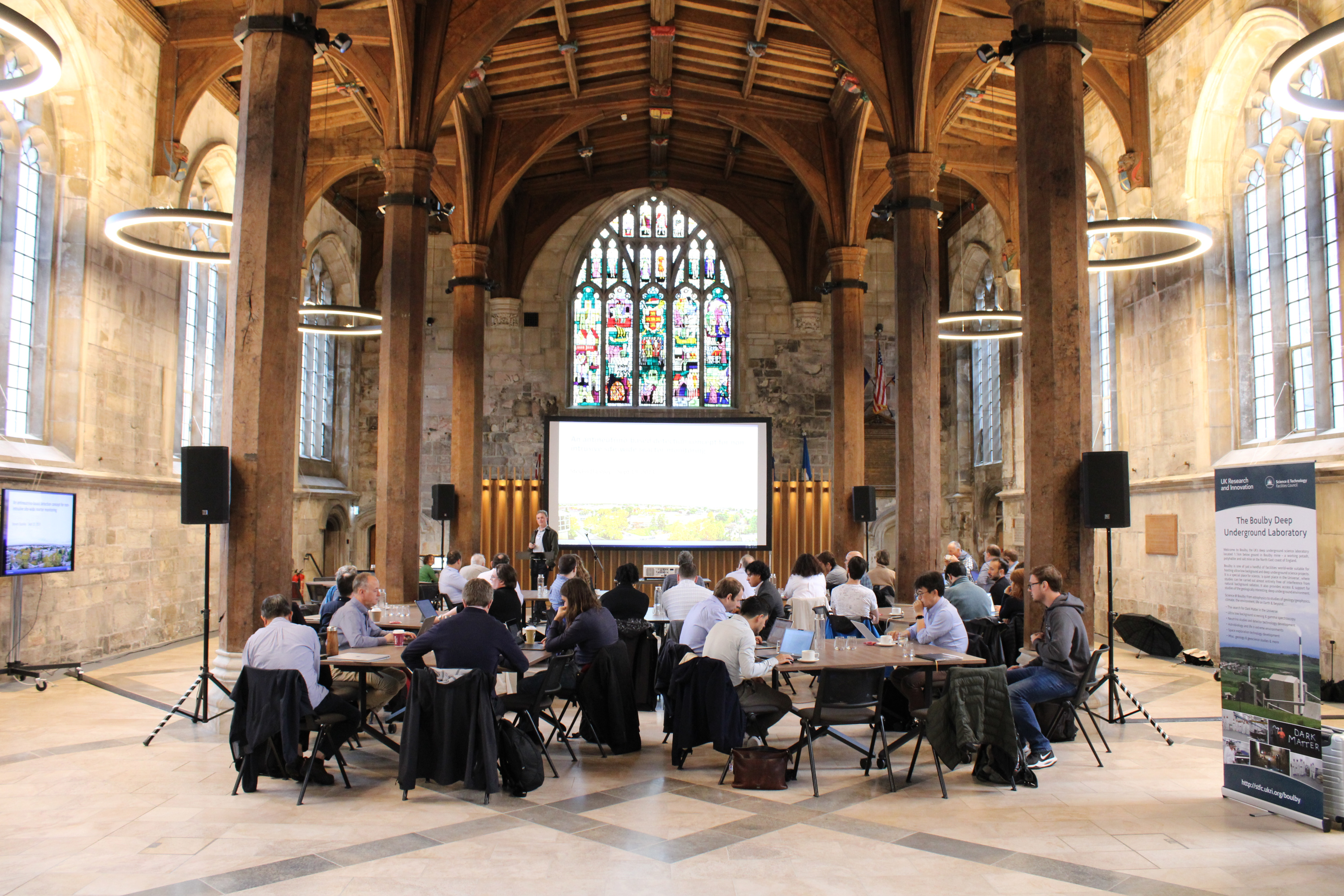}
    \caption{The lecture hall for the AAP 2023 workshop is the historic 
 15th century Main Hall of the Guildhall. The support columns are of solid oak. In this photograph, Steven Dazeley is about to deliver a lecture on ``Antineutrino detection and technology overview''.}
    \label{fig_DazeleyLecture}
\end{figure*}
 
In the opening speech, John Learned gave an overview of the workshop series, how they started and the current scope.
As the founder of the workshop series, he recalled a sad beginning in connection to the tragic 9/11/01 events. %
John's initial reaction, when asked how neutrinos could be used in nuclear security in light of those events, was that neutrinos were useless. Nevertheless, assuming an unlimited budget, he made estimates for the JASON committee and had some interactions with Freeman Dyson. The AAP workshop series was born.
The meetings over the last 20 years have been fruitful, and a primary venue for the discussion of neutrino applications.
The University of Hawai`i now hosts material from previous AAP workshops\footnote{\href{https://www.phys.hawaii.edu/aap/}{www.phys.hawaii.edu/aap/}}.

Sean Paling welcomed attendees on behalf of the co-host STFC Boulby Underground Laboratory, and Chris Toth gave an overview of the Laboratory, and the experiments currently ongoing in the deepest mine in the UK\footnote{\href{http://stfc.ukri.org/boulby}{stfc.ukri.org/boulby}}.
Following the effort to deploy AIT-NEO/WATCHMAN, there is now a plan to deploy a 30-ton BUTTON detector, as well as plans for BUTTON+ (a kt-scale detector). Boulby is situated at about 26 kilometers from the EDF Hartlepool nuclear power station. The dark-matter program was also highlighted as well  as muography-based tsunami early warning system and astrobiology experiments. The lab will see an expansion in the coming years, with support from the UKRI.
The lab, at about 1~km underground, provides 4,000~m$^2$ area of low-background environment, with a particularly low radon level of $\sim 3$ Bq/m$^3$.

There was an emphasis on new technology in the talk ``Antineutrino detection \& technology overview’’ by Steven Dazeley\footnote{The italic font indicates abstracts submitted by the speakers.}, who surveyed the technologies and techniques being developed now, and projected how they may improve capabilities for applications in the future.
{\it Since the successes of the Daya Bay, Double Chooz and RENO, $\theta_{13}$ experiments in the early 2000s, antineutrino detection technologies have continued to evolve, providing new capabilities such as next
generation scintillating materials, photon detectors and new dual phase TPC techniques. Some of these
technologies are being employed in detectors coming online now. Most recently, aboveground detection
of reactor antineutrinos, which relies upon the identification of cosmogenic fast neutrons to reduce the
most prevalent backgrounds, was accomplished using pulse shape sensitive scintillator or innovative new
designs that permit improved topological reconstruction of complex event structures. Speculating on
future developments, we can look forward to capabilities such as improved aboveground sensitivity, and
shallower deployment requirements in general. Further advances will come from order of magnitude
improvements in vertex resolution, higher photon detection efficiencies, more stable and less toxic
materials, and better particle ID. Separately, coherent scattering detection remains an elusive but
potentially attractive solution due to the relatively high interaction cross section. Many of the new
technologies described ... will have important implications for applications in the future. }

Rachel Carr (with input from Tomi Akindele) gave an overview of ``Neutrino Applications: Highlights from 46+ years of ideas’’. Rachel highlighted cooperative and non-cooperative reactor monitoring, spent fuel and reprocessing monitoring, and nuclear explosion monitoring as potential neutrino applications, before uncovering some interesting facts about neutrino communication including a patent on a ``Navigation system based on solar neutrino detection” and an article in the New York Times dating back to the 1970s. She finished with a review of submarine tracking and core verification using neutrinos and a discussion of what is needed to take neutrino applications from idea to adoption.

Extensive coverage of the history of modeling reactor antineutrino spectra was given in the talk by ``Antineutrino Detection \& Technology Overview’’ by Leendert Hayen. This gave an overview of the last decade or so, beginning with the birth of the Reactor Antineutrino Anomaly (RAA) and appearance of the 5-MeV reactor bump. Leendert reviewed progress in reactor antineutrino spectral prediction with a discussion of the conversion method and related large-scale shell model calculations of dominant forbidden transitions to partial mitigation of the reactor bump and new beta spectral measurements to potential resolution of the RAA. A discussion of the summation method covered the improvements following many dedicated TAGS (Total Absorption Gamma-ray Spectroscopy) campaigns. An overview of reactor antineutrino detection in the past decade reviewed results on the total reactor flux, spectral agreement and fuel evolution but emphasized that there is as yet no uniform picture.

In the afternoon, there were two lectures followed by shorter talks. Patrick Huber kicked off the afternoon with his talk ``Global Project Overview – where do we go from here?’’, which reviewed current and R\&D projects focused on neutrinos for nuclear safeguards, highlighting upcoming work by the Virginia Tech group. This covered CEvNS projects, global surface and underground water Cherenkov IBD detection R\&D and ocean deployed detection, before discussing practical constraints in terms of backgrounds,  proximity to a reactor, and statistical and systematic uncertainties.

In the talk ``Neutrinos for Nonproliferation: Safeguards and Advanced Reactors", Andrew Conant gave remarks on the IAEA safeguards and safeguarding activities and how neutrinos relate to current and future activities. Andrew concluded that neutrino detection is unlikely to replace traditional technologies, but rather complement them, however it does have potential application to advanced reactor types. With a focus on molten salt, high-temperature gas and sodium- and lead-cooled fast reactors, as well as small modular reactors and higher enrichment levels, there was a discussion of how advanced reactor technologies pose new safeguarding and safeguards challenges and the ways in which neutrinos can address those challenges in a way that traditional monitoring methods cannot.

There were then a series of shorter talks highlighting various use cases and recent studies on ``Neutrino Applications’’.
Michael Foxe gave an overview of the report ``Nu Tools: Exploring Practical Roles for Neutrinos in Nuclear Energy and Security'', in which he described the details of the Nu Tools study and methods, the Nu Tools Framework, and cross-cutting and use-case specific findings for the study.
{\it The Nu Tools study was developed to explore the potential roles for neutrino within nuclear energy and nuclear security.
This effort differs from previous neutrino detector studies as it is focused on the potential utilities and determining if there is
a possible use case for neutrino detectors as a monitoring technology. Due to the importance of understanding potential use
cases, this effort focused on interviewing experts working in the respective application areas. These experts focused on
nuclear safeguards and nuclear reactor operations, while also including a set of neutrino technology experts.}

In ``Nuclear Safeguards: Monitoring of Spent Nuclear Fuel'', Yan-Jie Schnellbach explored how spent nuclear fuel (SNF) can be monitored through the detection of antineutrinos from fission fragments present in the spent fuel. The sensitivity of idealised detectors with three different detection media to SNF in an underground geological repository and a surface interim storage facility was explored for different monitoring scenarios and SNF monitoring with CEvNS was also covered.

Andrew Conant presented results from a ``Sensitivity Tool for Antineutrino Monitoring of Small Modular Reactors''. {\it While reactor antineutrino detection has been performed at current commercial and research reactors, their detection at
advanced reactors poses new challenges. The NuTools study identified that neutrino detection could play a role in the
safeguards of advanced reactors, such as small modular reactors (SMRs). Several SMR concepts focus on the operation of
multiple modules, often within the same reactor building, to be flexible to load demand. This concept of operations would
likely require more intensive safeguards, e.g., more frequent inspection visits. We focus on the sensitivity of multiple
neutrino detectors at SMR sites. Cases of interest include total and individual reactor signals for operational status, power
level, and the possibility at fissile content determination. Results show that operational status is relatively simple to
determine for the sum of the modules but the sensitivity of discriminating all units on versus all but one of the units on to be
difficult depending on the proximity of the detector to the individual reactor unit. Power level quantification is still underway.}

``Establishing Antineutrino-based Safeguards Using the State-level Concept'' by Caiser Bravo (on behalf of Matthew Dunbrack, the GeorgiaTech group) discussed a methodology to propose potential use cases with plausible sensitivity goals for IAEA safeguards activities.
{\it Antineutrino detection systems have potential to safeguard the next generation of nuclear reactors. In theory, these systems
can be implemented for status verification, reactor power monitoring, and special nuclear material diversion detection.
Previous studies have investigated the applicability of this technology using models and simulations, but are often heavily
reliant on various situational assumptions, such as diversion pathways and detection probability limits. As
antineutrino-based detection systems continue to be considered for future nuclear reactor safeguards, researchers need a
coherent and flexible method to adjust system parameters to match the current International Atomic Energy Agency (IAEA)
framework.
In this work, we develop a methodology for selecting detection probability limits for antineutrino-based safeguards.
Detection probability, or the probability of detecting a diversion scenario, is a key metric for assessing current
antineutrino-based safeguard capabilities. Highlighting the IAEA's utilization of State-level safeguards approaches, our
detection probability threshold values are quantified by State-specific factors, or factors used by the IAEA in establishing
safeguards activities. We interpolate between the various State-specific factors and user-given weighting parameters to
quantify reasonable detection probability limits. This detection probability can shift drastically depending on a wide-range of
situational parameters, including reactor and facility design.   The    results indicate that antineutrino-based safeguards would
perform best as a complementary safeguard, regardless of reactor design.}

``Scalability of Gd-doped water-Cherenkov reactor-antineutrino IBD detectors for non-proliferation'' by Viacheslav Li presented results from a study of real-world monitoring ranges as a function of detector size.
{\it Recent advances in large water-Cherenkov detector technology, such as doping water with gadolinium at
Super-Kamiokande, highlight the feasibility of detecting antineutrinos from power reactors hundreds of kilometers away.
However, in the context of nuclear non-proliferation, detecting power reactors is generally not considered a challenge using
more traditional detection techniques. Of more interest are relatively small (50MWt) nuclear reactors which can potentially
evade detection.
The question is then: with an accurate understanding of detector efficiencies and backgrounds, how scalable is the
Gd-doped water-Cherenkov detection technique?
If we have a multi-kiloton detector, what is the maximum range to detect such a modest reactor in one year?   This talk reported on    a recent study of the scalability of Gd-doped water Cherenkov detectors in three different reactor antineutrino
background environments chosen to represent regions with high, medium and low concentrations of nearby large reactors.}

\section*{Day 2}

\subsection*{Neutrino applications: Session 2} %
The morning session was opened with the talk ``An antineutrino-based detection concept for non-intrusive site-wide reactor monitoring'' by Steven Dazeley. 
{\it 
Future advanced reactor designs may use liquid fuel or uncountable numbers of small fuel elements. Such designs will be
difficult or impossible to monitor for safeguards via traditional item accountancy techniques. One way to address this issue,
being pursued by the U.S. Dept of Energy, Office of Nuclear Energy Material Protection Accounting and Control
Technologies (MPACT), is to develop instruments that can operate directly inside the core. To this end, diagnostic tools are
being developed that can withstand the extreme temperature and radiation conditions near the fuel inside a reactor.
Generally, however, due to the harsh conditions, compromises to sensitivity must be considered in favor of long-term
survivability. Another approach is to monitor from a safe distance is via antineutrino detection. Above ground tools are being
developed to monitor antineutrino flux (PROSPECT, miniCHANDLER and MAD). However, these tools must be placed
where the reactor flux is high enough to overcome the high rate of cosmogenic backgrounds present above ground. For low
flux scenarios such as low power reactor monitoring, or in situations where infrastructure around the reactor does not permit
a close in deployment, larger liquid-based detectors deployed a few meters underground may be required to reduce the
hadronic component of the cosmogenic background. For detector target volumes greater than about $\sim$10 tons, it is worth
considering simple monolithic detector designs that can be deployed quickly. By reducing channel count and making use of
novel scintillator formulations that can improve background rejection, it may be possible to improve deployability and
decrease overburden compared to the state-of-the-art monolithic detectors, such as Double-Chooz and Daya Bay
detectors... The results of a MC-based analysis of a simple design that can provide from between 10-100
tons of fiducial target, while limiting overall non-fiducial detector size   were presented, and    indicate that the chosen design can deliver
excellent background suppression and energy resolution.}

A thought-provoking talk ``SNIFR - Submarine Neutrino Identification For Reconnaissance'' was given by Alex Goldsack.
{\it 
We are now in an era where the sensitivity and scalability of neutrino detectors allows for more intimate investigation into
nuclear reactors at range. The reactors deployed on nuclear submarines have a thermal power output around an order of
magnitude lower than that of power reactors, but a scalable detector technology could be deployed on a large commercial
ship to detect submarines at a range comparable to passive sonar in some scenarios, independent of environment.   This talk presented    the viability of this technique in the context of modern detector sensitivities, and discussed its practicality.
}

\subsection*{Neutrino detection \& technology: Session 1} %
``WbLS with pulse-shape discrimination'' by Tomi Akindele covered recent advances in the synthesis of water-based liquid scintillator (WbLS) by the LLNL team. Tomi presented a new formulation of high-concentration WbLS with pulse-shape discrimination for particle identification. Ongoing work on characterisation of formulations and on lithium doping of WbLS formulations for neutron detection was discussed.

In his talk on ``Materials development in the 30-tonne tank at BNL'', Minfang Yeh discussed the development of water-like WbLS (1-10\% LS) and oil-like WbLS ($>$90\% LS) and the potential for  and covered the scaling up of a bench-top WbLS experiment which investigated a variety of WbLS formulas and demonstrated doping with Gd, Li and B. Progress at a ton-scale  testbed at Brookhaven National Laboratory and the development of a filtration system in a 30-ton testbed were presented.

Jon Coleman gave a talk on the ``BUTTON technology testbeds'' planned for construction in the ultra-low background environment at STFC Boulby Underground Laboratory. The first of these - BUTTON 30 - will feature a modular support structure for testing advanced photosensors, and materials compatibility is being designed into the detector in anticipation of fills with advanced detection media such as Gd-doped WbLS.

\subsection*{Neutrino detection \& technology: Session 2} %
The talk on ``Measurement of reactor neutrinos using plastic scintillator cube'' by Shoichi Hasegawa covered the status and plan of neutrino monitor experiments at research reactors in Japan. 
{\it 
A new reactor neutrinos detector using a plastic scintillator is developed. This detector is compact and intended to measure
$\bar{\nu}_{e}$ from nuclear reactors by ground-based installation. The compact one-ton class detector for inverse beta
decay (IBD) must be installed close to the $\bar{\nu}_{e}$ source. For this purpose, the detector is being developed near the
core of a research reactor. For ground-based detectors, it is an important issue to distinguish the $\bar{\nu}_{e}$ signals
from background events. In this study, a plastic scintillator cubic detector with high position resolution will be developed to
improve the background event rejection in the prompt of IBD signal.
}

Sertac Ozturk gave a talk on ``Nuclear reactor monitoring with gadolinium-loaded plastic scintillator modules'' .
{\it 
In this talk, simulation-based design and optimization studies of a gadolinium-loaded segmented plastic scintillator detector
were presented for monitoring applications of nuclear reactors in Turkey using antineutrinos. Synthesis and characterization
results of gadolinium-loaded plastic scintillator samples were discussed.
}

Cristian Roca talked about ``A $^6$Li-doped pulse shape sensitive plastic scintillator for ton-scale detector applications''.
{\it 
Large-scale $^6$Li-doped pulse shape sensitive plastic scintillator is one of several technologies under development within the
Mobile Antineutrino Demonstrator project. Liquid scintillator with similar capabilities was one of key aspects of the
aboveground reactor antineutrino detection demonstration by the PROSPECT experiment. However, a plastic material is
considered a requirement for truly mobile above-ground detection systems suited to reactor monitoring for safeguards. The
new formulation of plastic scintillator is being developed in partnership with Eljen Technologies and can be obtained in
multi-liter single volumes enabling the construction of segments at meter-scale lengths.   A    summary of
measured performance criteria, which include attenuation length, stability, pulse shape sensitivity, and neutron efficiency
measurements   was presented].
}

``Performance of the ROADSTR PSD Plastic Prototype Detector''
was discussed by 
Caiser Bravo.
{\it
The Reactor Operations Antineutrino Detection Surface Testbed Rover (ROADSTR) detector prototype was constructed
from Li-6-doped Pulse Shape Discriminating plastic scintillators. Comprising 36 bars of 50 cm in length arranged in a 6$\times$6
array, ROADSTR has a mass of about 60~kg. Over almost one year of operation, this device has been used to study
scintillator characteristics and inverse beta decay backgrounds. The characterization of the detector
and a variety of background measurements conducted with it were described.
}

\subsection*{Neutrino detection \& technology: Session 3} %

A talk on ``Novel opaque scintillator technology for antineutrino detection" was presented by Thiago Bezerra.
{\it Scintillator detectors have been used for antineutrino detection since the 1950s when Cowan and Reines used them to
discover the neutrino. Modern experiments still use scintillators to study neutrino physics. Scintillators convert the energy
released in a neutrino interaction into light, which photosensors can detect. Traditional scintillator detectors are transparent,
allowing light to reach the photosensors. However, this transparency also limits the ability to image the neutrino interaction.
Scintillator detectors can be segmented to improve their imaging, which introduces additional challenges in building and
operating the detector. This talk detailed LiquidO, a new and counterintuitive opaque scintillator detector. Opacity is achieved  with short scattering length materials for the scintillation light. The opaque scintillator is traversed by
wavelength-shifting fibres, which collect and transport the light to the photosensors. This arrangement allows for
high-resolution imaging, enabling highly efficient particle identification from the MeV to GeV scale and many applications.
It was shown  how LiquidO can discriminate signal from backgrounds with high significance when measuring reactor
antineutrinos at the surface without an underground facility. Finally, prototype results demonstrating the
LiquidO technique and the plans for constructing a 5-ton demonstrator close to France's Chooz nuclear power plant were presented.}

Anatael Cabrera was unable to attend to give his talk about a ``Novel Methodology for Low Energy IBD-like Antineutrinos Detection and Potential''.
{\it The novel methodology outlined in the just-released “Probing Earth's Missing Potassium using the Unique Antimatter
Signature of Geoneutrinos” (\href{https://arxiv.org/abs/2308.04154}{arXiv:2308.04154}) shows a novel possible way to address one of the most extreme
measurements neutrinos may be able to accomplish: the observation of $^{40}$K geoneutrinos — so far proved impossible. The
most challenging condition of this measurement is to find an IBD-like interaction with an energy threshold well below the
typical 1.8MeV IBD on protons since the energy spectrum of $^{40}$K has a Q-value of ~1.3MeV. Our study finds, for the first
time, that there seems to be only a single isotope in the Universe capable of addressing this elusive requirement as well as
others necessary to ensure feasible detection. One of the most critical is that detection requires a detector capable of
identifying the signature of antimatter; i.e. single-e+ ID is a must. This rules out most of today’s technology except the
LiquidO (see dedicated talk), which may be able to address the challenges upon loading. Beyond geoneutrino observations,
there are possible applications that open with lower-energy anti-neutrinos detection using the same principle.}

Igor Jovanovic gave a talk on the ``Development of a High-Energy Two-Component Gamma Calibration Source".
{\it The detection of electron antineutrinos can provide the means for confirming the presence and monitoring of the operational
characteristics of nuclear reactors. Water-based Cherenkov detectors with gadolinium doping are one of the technologies
under study for this application. The energy scale of the emitted positron and the de-excitation cascade from neutron
capture by gadolinium motivates the development of gamma calibration sources with energies of several MeV. One such
potential source is provided by the $^{13}$C($\alpha$,n)$^{16}$O reaction. At alpha energies above $\sim$5 MeV, a significant branching ratio
exists for the deexcitation of 16O via the emission of a 6.1-MeV gamma ray. The fast neutron also produced from this
reaction can be used to tag events in the large water-based detector. 241Am is an appealing alpha source as it has
emission energy above the ~5 MeV threshold, high specific activity, and does not possess the regulatory overhead of other
alpha sources. Continued refinement of the simulation methods developed to predict source yield as a function
of the source design parameters   were discussed]. These include implementing more advanced physics models and transitioning the
simulation software to a more generalized framework. Initial measurement results to demonstrate
the production of the calibration signals of interest were additionally presented.}

John Learned gave a presentation on a ``Forest of Tubes for a directional IBD detector".
FoRest Of Scintillating Tubes (FROST) is a new concept to better identify directionality in inverse beta decay. Separated self-contained subunits (trees) are filled with liquid scintillator and have a dual ended PMT or SiPM readout. The studies at the University of Hawai`i are underway to explore various fills, light collection, and geometry to better assess the viability of the project.

``EoS – A Pathfinder Experiment for Low Energy Neutrino Physics with the Hybrid Detector THEIA" was presented by Hans Steiger.
{\it Future ktonne-scale, scintillation-based neutrino detectors, such as THEIA, plan to exploit new
and yet to be developed technologies to simultaneously measure Cherenkov and scintillation
signals in order to provide a rich and broad physics program. These hybrid detectors will be
based on fast timing photodetectors, novel target materials, such as water-based liquid
scintillator (WbLS), and spectral sorting. Besides a brief overview on THEIA’s program for
low energy astroparticle and particle physics this talk   focused    on a currently realized
demonstrator experiment, called EOS. This novel detector with an approximately 4-tonne target
fiducial volume is under construction at the UC Berkeley and LBNL (Lawrence Berkeley
National Laboratory). The detector will provide a test-bed for these emerging technologies
required for hybrid Cherenkov/Scintillation detectors. Furthermore, EOS will deploy
calibration sources to verify the optical models of WbLS and other liquid scintillators with slow
light emission, to enable an extrapolation to ktonne-scale detectors. This input will support the
development of advanced techniques for reconstructing event energy, position, and direction in
hybrid detectors significantly. After achieving these goals, EOS can be moved near a nuclear
reactor or in a particle test-beam to demonstrate neutrino event reconstruction or detailed event
characterization within these novel detectors.}

Patrick Huber presented a brief update on behalf of the PALEOCCENE Collaboration. He highlighted the potential application of CEvNS detection as a passive neutrino detection for reactor monitoring as discussed in the recent publication \href{http://arxiv.org/abs/2104.13926}{arXiv:2104.13926}.

\subsection*{Flux and spectrum prediction: Session 1} %

``Geoneutrinos: messengers from the inaccessible Earth" was presented by Virginia Strati. {\it The 99\% of the Earth’s radiogenic heat is generated by K, Th, and U that through beta minus decay release antineutrinos
and heat proportionally. The U and Th geoneutrino flux measured by underground liquid scintillator detectors aids in testing
Earth's compositional models with the energy spectrum analysis limiting U and Th quantity and distribution in the whole
planet. Accurate predictions of lithospheric geoneutrino signals, derived by constructing geophysical and geochemical
models, permit to improve the understanding of direct geoneutrino measurements revealing mantle's radiogenic power and
composition.
Future years will see the geoneutrino data expand beyond the Borexino and KamLAND experiments. The imminent release
of data from the Canadian SNO+ experiment, along with the nearly complete Jiangmen Underground Neutrino Observatory
(JUNO), points out a new age of multi-site geoneutrino detection, enhancing our comprehension of geoneutrino signals
originating from the Earth.
The talk   reviewed    the impacts of the recent results from KamLAND and Borexino, the expected outcomes from SNO+ and
JUNO and the future perspectives and challenges for geoneutrino science.}

Bedrich Roskovec's talk on ``Reactor flux from reactor data" asserted that the historical transition from total flux measurement to isotopic flux measurement in the context of reactor antineutrinos reflects a significant advancement in experimental techniques aimed at extracting complex information about neutrino flux and energy spectra. Initially, total flux measurements provided a general understanding of neutrino emissions, but recent experiments, such as those discussed by Cristian Roca (see below), have focused on measuring flux and spectrum across specific energy bins. At DayaBay, the evolution of fuel over these energy bins revealed that the measured spectrum, representing the average yield in each bin, was inconsistent with both the Huber-Mueller (HM) and Summation Model (SM2018) predictions. Notably, the evolution slope of electron antineutrinos for the SM2018 model showed a small tension with the HM model, highlighting the complexities and nuances in understanding reactor antineutrino behavior.

Christian Roca gave a talk on ``Reactor spectrum from reactor data".
In the past decade, significant global efforts have focused on solving the reactor antineutrino anomaly and measuring antineutrino spectrum with high precision, leading to remarkable advancements. High-precision measurements have revealed shortcomings in existing models, particularly highlighting a "bump" anomaly in the spectrum. As more data has been collected, researchers are increasingly identifying the isotopes responsible for this anomaly and quantifying their contributions. Recent initiatives to analyze underlying nuclear data and refine predictions are further clarifying the causes and potential solutions. Additionally, international collaboration is establishing a comprehensive database of spectra that will serve as a benchmark for future studies. This talk   highlighted    recent measurements done by DoubleChooz, DayaBay, NEOS-II, RENO, STEREO, and PROSPECT (as well as joint analysis done by DayaBay, STEREO, and PROSPECT).

Alejandro Sonzogni talked about the ``Progress towards understanding the source of the Reactor Antineutrino Anomaly" A review of 
{\it the nuclear data used in the normalization of the electron spectra measured at the Institut Laue Langevin
in the 1980s, concluding that they are very close to currently recommended values, except for the neutron capture cross
section on 207Pb, which is 9\% higher. This would lead to an artificially larger $^{235}$U electron and antineutrino spectra,
consistent with the DayaBay Collaboration results, as well as those reported recently by Kopeikin and collaborators.
Additionally, following an analysis that employs the latest nuclear databases of the electron data measured at ORNL in the
1970s by Dickens and collaborators,   they    have deduced new electron and antineutrino spectra for $^{235}$U and $^{239}$,$^{241}$Pu
under equilibrium conditions, which are consistent with the above mentioned normalization issue, and which can better
reproduce the IBD antineutrino spectrum near its maximum, thus providing a coherent explanation for the origin of the
Reactor Antineutrino Anomaly.}

A talk on the ``Fission yields of isomers in antineutrino calculations" was given by Andrea Mattera.
{\it Isomeric states have been observed in about 150 of the hundreds of isotopes that can be produced in the fission of major
actinides. These isomers can be populated directly through fission, and the isomeric yield ratio (IYR) represents the relative
population of the excited state(s) and the ground state (GS) independent yield.
In this work, we present a comprehensive study of the extent to which IYRs affect the antineutrino flux predictions with the
summation method using two different approaches. First, we estimated how a set of newly evaluated recommended IYRs
change the antineutrino spectra of all major actinides of interest for reactor antineutrino spectra
($^{235,238}$U,$^{239,241}$Pu). Then we individually looked at the contribution of each fission product with a known
isomer, and studied how a different IYR value would affect the calculated antineutrino spectra.
While essentially no effect on the antineutrino spectrum is observed below 5 MeV, changes on the order of 1\%-2\% for each
fuel type become evident between 5 and 7 MeV. These grow to as much as 30\% above 7 MeV. The changes show
consistently an increase in the antineutrino yield when the newly evaluated isomeric yields are used, compared to the
values included in evaluated Fission Yields libraries.}

Paraskevi (Vivan) Dimitriou summarised the recent IAEA Technical Meeting on Nuclear Data for Reactor Antineutrinos and Applications. She showed recent highlights from reactor antineutrino experiments, antineutrino applications, flux and spectrum modelling, nuclear data and data preservation and dissemination. The final recommendation of the meeting was to form an international Working Group under the auspices of the IAEA in order to coordinate efforts and mitigate issues with limited resources.

Lorenzo Perisse was not able to attend in person but submitted a recording of their presentation on revising the summation calculation of reactor antineutrino spectra with the NE$\nu$FAR project.

The day was rounded off with a conference dinner with drinks reception courtesy of John Caunt Scientific Ltd at York's historic Hospitium. 

\section*{Day 3} 

\subsection*{Flux and spectrum prediction: Session 2} %

Andrew Petts gave a talk on ``Modelling of the anti-neutrino emissions from an Advanced Gas-cooled Reactor." {\it The Hartlepool Advanced-Gas-cooled Reactor (AGR) design   was    described and results from simulations of the
anti-neutrino emissions from the reactors based on actual in-core data presented. The potential for siting small antineutrino
detectors on site and performing near-field measurements and stand-off measurements   was also    discussed.}

The ``Anti-Neutrino Flux from the EdF Hartlepool Nuclear Power Plant" was discussed by Robert Mills. This talk presented {\it the first detailed simulation of the antineutrino emissions from an Advanced Gas-cooled Reactor (AGR) core,
based upon operational data from the UK Hartlepool reactors and reactor calculations for each of the 2592 assemblies in
each of the two cores. An accurate description of the evolution of the anti-neutrino spectrum of reactor cores is needed to
assess the performance of antineutrino-based monitoring concepts for non-proliferation, including estimations of the
sensitivity of the antineutrino rate and spectrum to fuel content and reactor thermal power. The antineutrino spectral
variation presented, while specific to AGRs, helps provide insight into the likely behaviour of other reactor designs that use
a similar batch refuelling approach, such as those used in RBMK, CANDU and other reactors. Comparisons were shown
with PWR reactor anti-neutrino emissions and the effects of different refuelling approaches.}

In the talk on ``Estimating Neutrino Signals Using \href{http://reactors.geoneutrinos.org}{reactors.geoneutrinos.org}", Steve Dye gave an {\it interactive tour of the neutrino signal estimates available using the online tool at reactors.geoneutrinos.org.
Questions, feedback and recommendations from participants were encouraged.}

``CONFLUX - The Reactor Antientrino Flux Prediction Software" (\href{https://github.com/CNFLUX/conflux}{github.com/CNFLUX/conflux})
was presented by Xianyi Zhang.
{\it The predicted reactor antineutrino flux is an important ingredient for particle physics measurements and neutrino-based
safeguards applications, ranging from neutrino oscillation measurements to monitoring reactor fuel and operations. Over the
past decade, comparisons between predictions and reactor neutrino experiments have revealed significant discrepancies
which have motivated new neutrino and nuclear data measurements. CONFLUX, the Calculation Of Neutrino FLUX, is a
software framework that aims to provide a flexible and modular tool for multiple communities. This new framework is being
developed to standardize the input and output of the neutrino flux calculation, increase the accessibility of neutrino, nuclear
data to the community, and package benchmark reactor and nuclear data. The software integrates three different prediction
modes: summation, beta-spectra conversion, and direct neutrino measurements. The comprehensive and flexible inclusion
of nuclear data allow users to perform sensitivity studies, evaluate impact of new data, monitoring studies, assess novel
reactor types, etc. In this presentation... the status of the framework development, the calculation capability, and
the potential applications   were described].}

\subsection*{Global projects: Session 1} %

The latest on ``Antineutrino measurements in SNO+" was presented by Sofia Andringa.
{\it SNO+ is a multi-purpose experiment located at SNOLAB in Canada, with the main goal to search for neutrinoless double
beta decay but measuring also neutrinos from several sources. In an earlier phase, SNO+ has made the first ever
observation of reactor antineutrinos in a pure water Cherenkov detector. Presently, the water has been replaced by liquid
scintillator, making SNO+ sensitive to geoneutrinos and allowing for much more detailed measurements of the reactor
antineutrino energy spectrum. Most of the flux comes from Ontario's nuclear power plants, located 250 km and 340 km
away from the detector, which provides a good sensitivity to neutrino oscillation parameters. This contribution   reviewed    the
past results, present status and future prospects for antineutrino measurements at SNO+.}

Lucas Machado gave an update on the status of reactor neutrinos in Super-Kamiokande Gadolinium. He presented the results of the first reactor antineutrino search in SK-Gd and prospects for future searches.

Thiago Bezerra gave a talk on the ``DoubleChooz Experiment: Latest Results for Antineutrino Applied Detection". {\it The latest results of the DoubleChooz experiment in the context of applied antineutrino detection were summarised in this
talk. DoubleChooz experimental set-up, based at the EDF Chooz nuclear reactor (France), offers full reactor power
modulation data. This enables DoubleChooz to achieve today’s world precision on the reactor flux measurement using
reactor-on data and a precise spectral characterisation with reactor-off data.}

An update on the PROSPECT experiment was give by Bryce Littlejohn.
PROSPECT has achieved a significant milestone by demonstrating over one signal-to-background ratio in an overburden-free reactor inverse beta decay (IBD) experiment, marking a major accomplishment for the Applied Antineutrino Physics (AAP) initiative. Throughout this process, the team has developed various technologies, tools, and knowledge, including leading sterile oscillation limits and reference 235U spectra, lithium-doped PSD-capable liquid scintillators, and innovative IBD detector design concepts. Additionally, they have created versatile and reliable cosmic background simulations and established a user-friendly reactor neutrino lab based in the U.S. at HFIR. The team is also progressing towards the multi-site deployment of PROSPECT-II. New directionality results were also highlighted in this talk.

Yury Shitov presented ``DANSS reactor antineutrino spectrometer: results for 2023".
{\it DANSS is a highly segmented solid-state scintillation spectrometer that detects up to 5000 reactor antineutrinos at a
distance of 10-13 m from the industrial nuclear reactor (4 units, 3.1 GW) of the Kalinin NPP. Taking into account the low
background (only 2\%), this makes it possible to search for oscillations into sterile neutrinos in the $\Delta m^2 \in (0.02,\; 5.0)\; \mathrm{eV}^2$
region. Results were presented based on record statistics --- almost 8 million reactor antineutrinos over 7 years of
measurements. They did not find a statistically significant oscillation signal and excluded a significant part of their possible
phase space. In addition, the results of research into the evolution of nuclear fuel, differences in the shapes of the
experimental and theoretical spectra of reactor antineutrinos were presented. Particular attention was paid to the
upgrade of the spectrometer, the main goal of which is to achieve an energy resolution no worse than 12\% at 1~MeV on a 1.7 times
more sensitive volume with the same passive shielding and mobile platform.}

\subsection*{Global projects: Session 2} %

An update on the ``Status of miniCHANDLER and CHANDLER" was given by Keegan Walkup.
CHANDLER utilizes a 3D optical lattice technology to effectively identify inverse beta decays (IBDs), distinguishing positrons through topology and neutrons via a long-decay scintillator. This technology has been experimentally validated, and moving from the prototype stage to a full-size detector presents opportunities for enhancements in optics, engineering, electronics, and analysis tools. Additionally, a CHANDLER detector is set to be deployed as part of the Mobile Antineutrino Demonstrator. Updates on 3D reconstruction, topology, optics, and electronics were presented in this talk.

Carl Metelko gave a talk on the ``Measurement of anti-neutrinos from Nuclear Waste with VIDARR." {\it VIDARR (Verification Instrument for the Direct Assay of Radiation at Range) is a ~2 tonne plastic scintillator detector doped with gadolinium designed to measure the anti-neutrino flux from reactor cores at a safe distance outside the core. The detector will shortly be deployed at Sellafield to detect anti-neutrinos from nuclear waste. A description of the detector and deployment   was    given.}

The ``Mobile Antineutrino Demonstrator (MAD) experiment." was presented by Nathaniel Bowden
Recent advancements in technology and a deeper understanding of neutrino applications have sparked a new initiative to create a mobile detection system. The MAD project is informed by Nu Tools, integrating feedback from potential users and hosts into its design. A simulation framework is being developed to predict detector performance, guide design, and validate background predictions. Two complementary plastic-based detection concepts with promising performance have been identified, and MAD will incorporate subsystems from both to enhance these technologies. Ultimately, MAD aims to deliver a Mobile Demonstrator system along with foundational technology and knowledge applicable to various near-field use cases.

Byeongsu Yang presented on ``The status of Reactor Experiment for Neutrino and Exotics".
{\it   They reported    on the status of Reactor Experiment for Neutrino and Exotics (RENE),
which primarily aims to search for the sterile neutrino oscillation at $\Delta m_{41}^2 \sim 2 eV^2$. The joint study of RENO and NEOS experiments showed a hint for the sterile neutrinos at $\Delta m_{41}^2 \sim 2.4 eV^2$ and $\sim 1.7 eV^2$, which overlap with the allowed region by the Reactor Anti-neutrino Anomaly. This experiment can also take precise measurements of the flux and spectrum of reactor electron antineutrino, and separate the reactor neutrino spectrum into those from 235U and 239Pu. In this presentation, the detector concept, physics cases, and the status of the experiment   was reported].}

\subsection*{Global projects: Session 3} %

Jason Newby gave a talk on ``The COHERENT Experiment at the Spallation Neutron Source"
{\it The first observations of coherent elastic neutrino nuclear scattering (CEvNS) on multiple nuclei were recently made by the COHERENT experiment at the Spallation Neutron Source at the Oak Ridge National Laboratory. This basic interaction now
lays the foundation for a new era in developing compact neutrino detectors as well as a new probe of physics topics
including electromagnetic properties, searches for physics beyond the standard model, and nuclear form factors. The
Spallation Neutron Source is ideally suited for not only CEvNS studies but also a broader set of high-precision neutrino
physics measurements and dark matter searches due to the accelerator's intensity, pulsed structure, and proton-beam
energy... An overview of the compelling scientific opportunities in particle physics enabled by proton power
upgrades now underway and a future upgrade of a new target facility at the SNS were presented.}

``Recent results from the CONUS experiment" were presented by Christian Buck.
{\it The CONUS reactor antineutrino experiment studies coherent elastic neutrino nucleus scattering (CEvNS) on germanium
nuclei. For several years, the experiment was collecting data at about 17 m distance from the 3.9 GWth reactor core of the
nuclear power plant in Brokdorf, Germany. Very low energy thresholds of about 210 eV were achieved in four 1 kg point
contact germanium detectors equipped with electric cryocooling. With the most recent data set, the constraints on the
CEvNS rate could be significantly improved as compared to previous CONUS analyses. The CONUS setup was recently
moved from Brokdorf to a power plant in Leibstadt, Switzerland, where the experiment will continue data taking with
improved detectors and an optimised shield design.}

Mark Lewis talked about ``CEvNS at the Dresden-II reactor and beyond".
{\it The recent detection of coherent elastic neutrino-nucleus scattering (CEvNS) enables neutrino investigations of new
physics with small-size detectors. However, CEvNS generates signals at the few- or sub-keV levels, requiring very sensitive
detector technologies. High-yield neutrino sources, including power reactors, provide the fluxes required for definitive
explorations of the phenomenological and technological applications of CEvNS.
The applicability of p-type point contact Ge detectors to this sector of neutrino physics is well established. Several
improvements to the current epitome of Ge diodes, the NCC-1701 detector, are in progress for its re-deployment in the next
phase of reactor CEvNS exploration.
In this talk,   Mark presented    the success of NCC-1701 at the Dresden-II core in the USA, its operational challenges, and the
slew of improvements built into its upcoming installation at the Ringhals nuclear plant in Sweden. In the context of Ringhals... the feasibility of reactor power monitoring via CEvNS, even under present detector capabilities   was also discussed].}

Han Steiger gave a talk on the latest on the JUNO-TAO experiment.
{ \it The Taishan Antineutrino Observatory (TAO or JUNO-TAO) is a satellite detector for the Jiangmen Underground Neutrino
Observatory (JUNO). JUNO will use reactor antineutrinos at
a baseline of $\sim$53 km to probe the interference effects between the two atmospheric mass-squared differences, which are
sensitive to the sign of the mass ordering. Located near the Taishan-1 reactor, TAO independently measures the
antineutrino energy spectrum of the reactor with unprecedented energy resolution and by that uncovering its fine structure
for the first time. Beyond that, TAO is expected to make world-leading time-resolved measurements of the yield and energy
spectra of the main isotopes involved in the antineutrino emission of nuclear reactors. By that TAO will provide a unique
reference for other experiments and nuclear databases. The TAO experiment will realize a neutrino detection rate of about
2000 per day. In order to achieve its goals, TAO is relying on cutting-edge technology, both in photosensor and liquid
scintillator (LS) development which is expected to have an impact on future neutrino and Dark Matter detectors. In this talk,
the design of the TAO detector with special focus on its new detection technologies was introduced. In addition, an overview of the progress currently being made in the R\&D for photosensor and LS technology in
the frame of the TAO project were presented.}

A talk on the ``Prospects for geo-neutrinos and supernova neutrinos with JUNO" was presented by Pierre-Alexandr Petijean.
{\it The Jiangmen Underground Neutrino Observatory (JUNO) is a medium-baseline reactor neutrino experiment under
construction in China. It consists of a 20 kt liquid scintillator detector designed for neutrino physics. The main objective of
the experiment is to determine the neutrino mass ordering by measuring the energy spectrum of reactor antineutrinos from
eight neighboring cores. Also, JUNO will be sensitive to neutrinos emitted by natural sources such as geo-neutrinos and
supernova neutrinos.
The measurement of the geo-neutrino provide information about the abundance of Uranium and Thorium in the Earth’s crust
and mantle, as well as our planet’s heat budget. Within the first year of data taking, JUNO will be able to exceed the
statistics on existing geo-neutrino flux results from Borexino and KamLAND experiments. With increased statistics, JUNO
will be able to measure Uranium and Thorium fluxes individually and to establish their ratio, giving insights about Earth's
formation processes.
In addition to geo-neutrinos, JUNO will allow measurements of the diffuse supernova neutrino background, pre-supernova
neutrinos and the all flavor neutrino flux from a Galactic core-collapse supernova (CCSN) with high statistics, low threshold
and high energy resolution. For maximizing the physics reach of JUNO as a neutrino telescope, two trigger systems will
operate to search for a transient astrophysical signal in real time: (i) a dedicated multi-messenger trigger system, (ii) a
prompt CCSN monitor embedded in the global trigger system. With those two systems, we will be able to collect precise
data that will help us understand the physics of CCSN and other astrophysical phenomena.
This talk presented the expected JUNO sensitivity to geo-neutrinos and the expected performance of JUNO for the
detection of different supernova neutrino fluxes.}

The AAP 2023 workshop was concluded with a summary and closing statements from Nathaniel Bowden. 

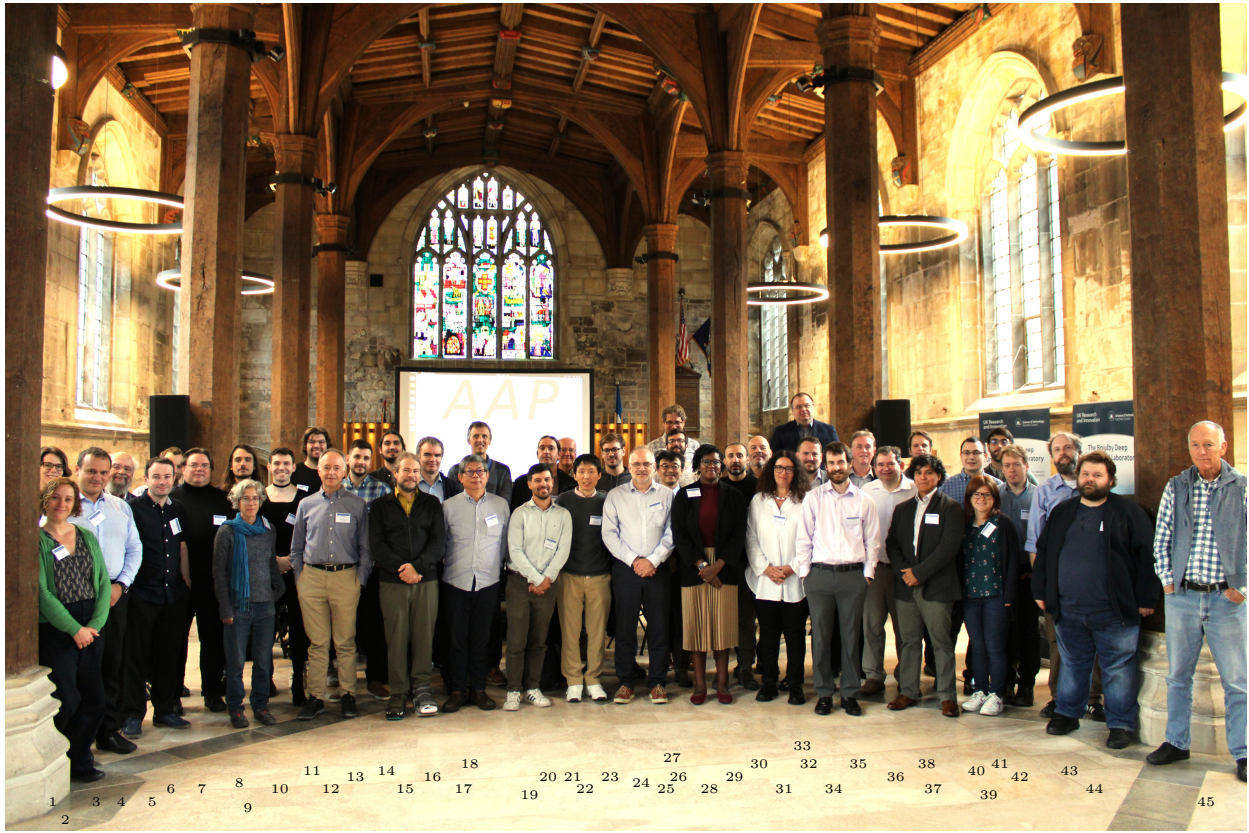
\begin{figure*}[ht]
\begin{overpic}[width=1.0\linewidth]%
{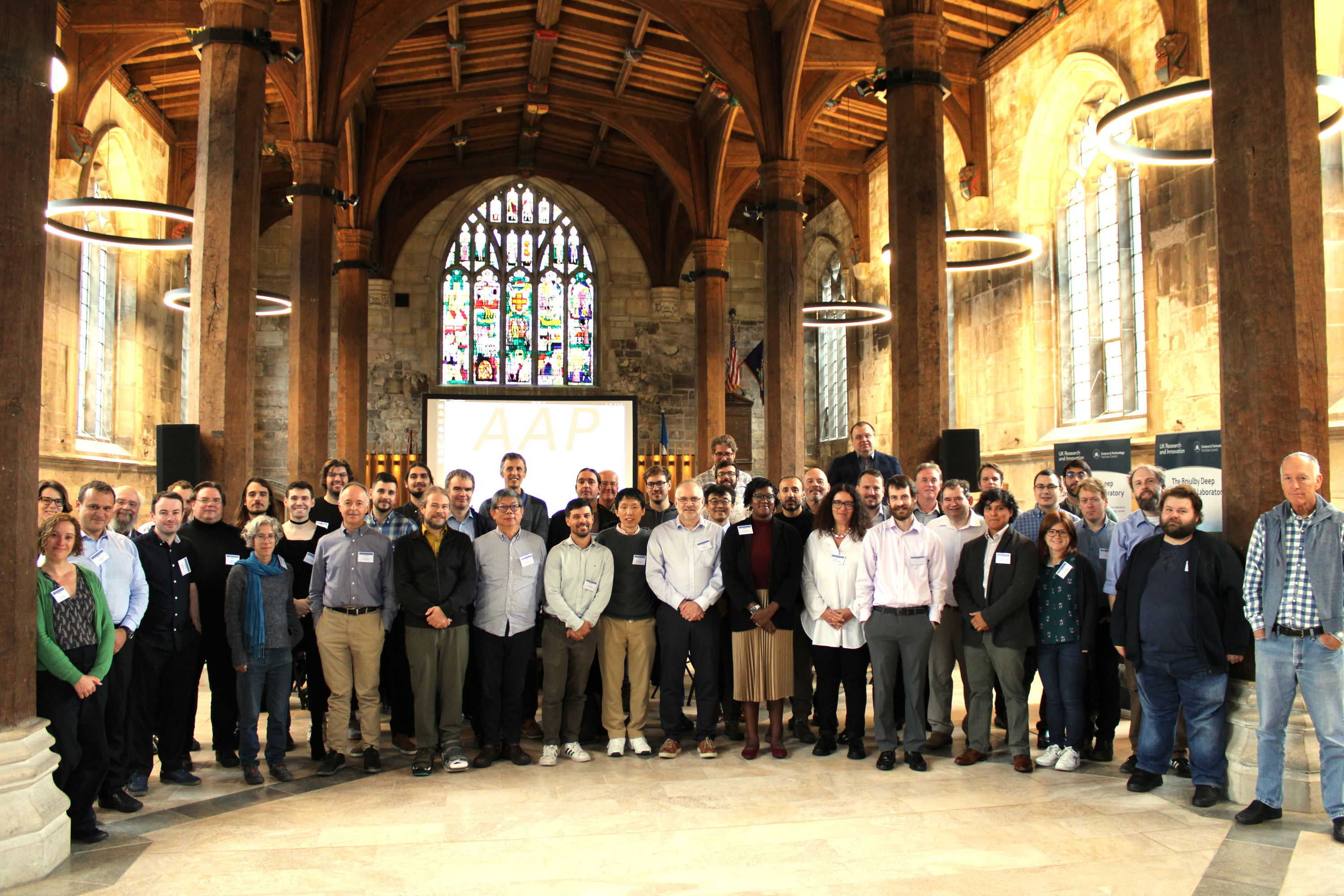}
{\tiny
{
\put(3.5,2){1}
\put(4.5,.5){2}
\put(7,2){3}
\put(9,2){4}
\put(11.5,2){5}
\put(13,3){6}
\put(15.5,3){7}
\put(18.5,3.5){8}
\put(19.2,1.5){9}
\put(21.4,3){10}
\put(24,4.5){11}
\put(25.5,3){12}
\put(27.5,4){13}
\put(30,4.5){14} 
\put(31.5,3){15}
\put(33.7,4){16}
\put(36.2,3){17}
\put(36.7,5){18}
\put(41.5,2.5){19}
\put(43,4){20}
\put(45,4){21}
\put(46,3){22}
\put(48,4){23}
\put(50.5,3.5){24}
\put(52.5,3){25} 
\put(53.5,4){26}
\put(53,5.5){27}
\put(56,3){28}
\put(58,4){29} 
\put(60,5){30}
\put(62,3){31}
\put(64,5){32}
\put(63.5,6.5){33}
\put(66,3){34}
\put(68,5){35}
\put(71,4){36}
\put(74,3){37}
\put(73.5,5){38}
\put(78.5,2.5){39}
\put(77.5,4.5){40}
\put(79.4,5){41}
\put(81,4){42}
\put(85,4.5){43}
\put(87,3){44}
\put(96,2){45}
}
}
\end{overpic}
    \caption{AAP-2023 participants:
Rachel Carr (1), 
Liz Kneale (2), 
Luigi Capponi (3),
John Learned (4),
Jon Burns (5),
Bedrich Roskovec (6),
Bjorn Seitz (7),
Cristian Roca (8),
Sofia Andringa (9),
Chris Toth (10),
Alex Goldsack (11),
Igor Jovanovic (12),
Lucas Machado (13),
Yan-Jie Schnellbach (14),
Patrick Huber (15),
Nathaniel Bowden (16),
Minfang Yeh (17),
Jeff Hartwell (18),
Andrea Mattera (19),
Carl Metelko (20),
Steve Quillin (21),
Byeongsu Yang (22),
Mark Lewis (23),
Jon Link (24),
Shoichi Hasegawa (25),
Leendert Hayen (26),
Jason Newby (27),
Tomi Akindele (28),
Sertac Ozturk (29),
Alejandro Sonzogni (30),
Paraskevi Dimitriou (31),
Andrew Conant (32),
Yury Shitov (33),
Michael Foxe (34),
Steven Dazeley (35),
Robert Mills (36),
Caiser Bravo (37),
Thiago Bezerra (38),
Virginia Strati (39),
Keegan Walkup (40),
Pierre-Alexandre Petitjean (41),
Andrew Petts (42),
Jon Coleman (43),
Hans Steiger (44),
Steve Dye (45). Photo by Viacheslav~Li.
}
    \label{fig_AAP_participants}
\end{figure*}

\end{document}